\documentclass[11pt]{article}
\usepackage{epsfig,amsfonts,latexsym}
\usepackage{graphicx,color}
\usepackage{algorithm,algorithmic}
\usepackage{subfigure}
\newtheorem{theorem}{Theorem}[section]
\newtheorem{lemma}[theorem]{Lemma}

\newtheorem{proposition}[theorem]{Proposition}

\newtheorem{definition}[theorem]{Definition}

\newenvironment{proof}{\noindent {\bf Proof:}}{$\hfill\square$}

\newcommand\card[1]{\left| #1 \right|}
\newcommand\p{{\textsf{P}}}
\newcommand\np{{\textsf{NP}}}
\newcommand\dtime{{\textsf{DTIME}}}

\newcommand\set[1]{\left\{#1\right\}}
\newcommand{\pd}{{\mathcal{P}}}

\newcommand{\R}{{\Delta}}
\newcommand{\Opt}{{\texttt{Opt}}}

\begin{document}
\title{Approximation algorithms and hardness for domination with propagation}
\author{ \small Ashkan Aazami  \\ \small \texttt{aaazami@uwaterloo.ca}\\
\small Department of Combinatorics and Optimization \\\small
University of Waterloo \and \small Michael David Stilp
\\\small \texttt{mstilp3@gatech.edu}\\ \small School of Industrial and Systems Engineering\\ \small Georgia Institute of Technology}
%\author{ Ashkan Aazami, Michael David Stilp \\
%\texttt{\{aaazami,mdstilp\}@uwaterloo.ca}\\
%Department of Combinatorics and Optimization \\
%University of Waterloo}
\date{\today}
\maketitle
\begin{abstract}
The power dominating set (PDS) problem is the following extension of
the well-known dominating set problem: find a smallest-size set of
nodes $S$ that power dominates all the nodes, where a node $v$ is
power dominated if (1) $v$ is in $S$ or $v$ has a neighbor in $S$,
or (2) $v$ has a neighbor $w$ such that $w$ and all of its neighbors
except $v$ are power dominated. We show a hardness of approximation
threshold of $2^{\log^{1-\epsilon}{n}}$ in contrast to the
logarithmic hardness for the dominating set problem. We give an
$O(\sqrt{n})$ approximation algorithm for planar graphs, and show
that our methods cannot improve on this approximation guarantee.
Finally, we initiate the study of PDS on directed graphs, and show
the same hardness threshold of $2^{\log^{1-\epsilon}{n}}$ for
directed \emph{acyclic} graphs. Also we show that the directed PDS
problem can be solved optimally in linear time if the underlying
undirected graph has bounded tree-width.
\\

\\

\end{abstract}

\section{Introduction \label{Sec:Intro}}
A {dominating set} of an (undirected) graph $G=(V,E)$ is a set of
nodes $S$ such that every node in the graph is in $S$ or has a
neighbor in $S$.  The problem of finding a dominating set of minimum
size is an important problem that has been extensively studied,
especially in the last 20 years, see the books by Haynes et al.\
\cite{Bib:HAHESL98a,Bib:HAHESL98b}. The problem is \np-hard
\cite{Bib:GAJO79}, a simple greedy algorithm achieves a logarithmic
approximation guarantee\footnote{An approximation algorithm for a
(minimization) optimization problem means an algorithm that runs in
polynomial time and computes a solution whose cost is within a
guaranteed factor of the optimal cost; the {\em approximation
guarantee} is the worst-case ratio, over all inputs of a given size,
of the cost of the solution computed by the algorithm to the optimal
cost.} \cite{Bib:JO74}, and, modulo the $\p\not=\np$ conjecture, no
polynomial time algorithm gives a better approximation guarantee
\cite{Bib:LUYA94,Bib:FE98}.

Our focus is on an extension called the {\textsc{Power Dominating
Set}} (abbreviated as PDS) problem.  Power domination is defined by
two rules; the first rule is the same as the rule for the
\textsc{Dominating Set} problem, but the second rule allows a type
of indirect propagation. More precisely, given a set of nodes $S$,
the set of nodes that are {\em power dominated} by $S$, denoted
$\pd_S$, is obtained as follows.
\begin{itemize}
\item[(Rule~1)]
if node $v$ is in $S$, then $v$ and all of its neighbors are in
$\pd_S$;
\item[(Rule~2)] (propagation)
if node $v$ is in $\pd_S$, one of its neighbors $w$ is not in
$\pd_S$, and all other neighbors of $v$ are in $\pd_S$, then $w$ is
inserted into $\pd_S$.
\end{itemize}

The set $\pd_S$ is independent of the sequence in which nodes are
inserted  by Rule~2. Otherwise, there is a minimal counter example
with two maximal sequences of insertions and an ``earliest'' node
that occurs in one sequence but not the other; this is not possible.
The PDS problem is to find a node-set $S$ of minimum size that power
dominates all nodes (i.e., find $S\subseteq V$ with $\card{S}$
minimum such that $\pd_S=V$). We use $\Opt(G)$ to denote the size of
an optimal solution for the PDS problem for a graph $G$. Throughout,
we use $n$ to denote the number of nodes in the input graph.

\begin{figure}[!h]
\begin{center}
\label{Fig:Baker}
    \input{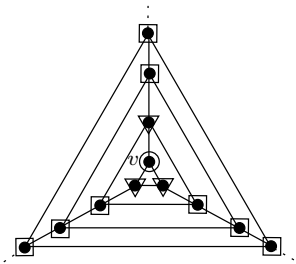}
 \caption{Illustrating those nodes power dominated by Rule 1 (denoted by a triangle) and Rule 2 (denoted by a square);
 the picked node is shown by a circle.}
\end{center}
\end{figure}

For example, consider the planar graph in Figure \ref{Fig:Baker};
the graph has $t$ disjoint triangles, and three (mutually disjoint)
paths such that each path has exactly one node from each triangle;
note that $|V|=3t$. The minimum dominating set has size
$\Theta(|V|)$, since the maximum degree is $4$. The minimum power
dominating set has size one -- if $S$ has any one node of the
innermost (first) triangle (like $v$), then $\pd_S=V$\footnote{In
more detail, we apply Rule~1 to see that all the nodes of the
innermost (first) triangle and one node of the second triangle are
in $\pd_S$; then by two applications of Rule~2 (to each of the nodes
in the first triangle not in $S$), we see that the other two nodes
of the second triangle are in $\pd_S$; then by three applications of
Rule~2 (to each of the nodes in the second triangle) we see that all
three nodes of the third triangle are in $\pd_S$; etc.}.

The PDS problem arose in the context of electric power networks,
where the aim is to monitor all of the network by placing a
minimum-size set of very expensive devices called phase measurement
units; these units have the capability of monitoring remote elements
via propagation (as in Rule~2); see Brueni \cite{Bib:BR93}, Baldwin
et al.\ \cite{Bib:BAMBA93}, and Mili et al.\ \cite{Bib:MIBP91}. In
the engineering literature, the problem is called the PMU placement
problem.

Our motivation comes from the area of approximation algorithms and
hardness results. The \textsc{Dominating Set} problem is a so-called
{\em covering problem}; we wish to cover all nodes of  the graph by
choosing as few node neighborhoods as possible. In fact, the
\textsc{Dominating Set} problem is a special case of the well-known
{\textsc{Set Covering}}\footnote{Given a family of sets on a
groundset, find the minimum number of sets whose union equals the
groundset.} problem.

Such covering problems have been extensively investigated. One of
the key positive results dates from the 1970's, when Johnson
\cite{Bib:JO74}, Lov\'{a}sz \cite{Bib:LO75} and later Chv\'{a}tal
\cite{Bib:CH79} showed that the greedy method achieves an
approximation guarantee of $O(\log|V|)$ where $|V|$ denotes the size
of the ground set, see also \cite{Bib:SL96}. Several negative
results (on the hardness of approximation) have been discovered over
the last few years: Lund and Yannakakis \cite{Bib:LUYA94} showed
that the \textsc{Set Covering} problem is hard to approximate within
a ratio of $\Omega(log{n})$ and later, Feige \cite{Bib:FE98} showed
that it is hard to approximate within a ratio of
$(1-\epsilon)\ln{n}$, modulo some variants of the $\p\not=\np$
assumption.

A natural question is what happens to covering problems (in the
setting of approximation algorithms and hardness results) when we
augment the covering rule  with a propagation rule. PDS seems to be
a key problem of this type, since it is obtained from the
\textsc{Dominating Set} problem by adding a simple propagation rule.
\subsection{Previous literature}
Apparently, the earliest publications on PDS are Brueni
\cite{Bib:BR93}, Baldwin et al.\ \cite{Bib:BAMBA93}, and Mili et
al.\ \cite{Bib:MIBP91}. Later, Haynes et al. \cite{Bib:HAHHH02}
showed that the problem is \np-complete even when the input graph is
bipartite; they presented a linear-time algorithm to solve PDS
optimally on trees. Kneis et al.\ \cite{Bib:KNMRR04} generalized
this result to a linear-time algorithm that finds an optimal
solution for graphs that have bounded tree-width, relying on earlier
results of Courcelle et al.~\cite{Bib:COMR98}. Kneis et al.\
\cite{Bib:KNMRR04} also showed that PDS is a generalization of the
\textsc{Dominating Set} problem as follows. Given a graph $G$ we can
construct an augmented graph $G'$ such that $S$ is an optimal
solution for the \textsc{Dominating Set} problem on $G$ if and only
if it is an optimal solution for PDS on $G'$; the graph $G'$ is
obtained from $G$ by adding a new node $v'$ for each node $v$ in $G$
and adding the edge $vv'$. Guo et al.\ \cite{Bib:GUNR05} developed a
combinatorial algorithm based on dynamic-programming for optimally
solving PDS on graphs of tree-width $k$. The running time of their
algorithm is $O(c^{k^2}\cdot n)$ where $c$ is a constant.   Guo et
al.\ also compared the tractability of the \textsc{Dominating Set}
problem versus PDS on several classes of graphs, that is, they study
whether there are classes of graphs where the former problem is in
\p\ but the latter one is \np-hard; but they have no result that
``separates'' the two problems. Even for planar graphs, the
{\textsc{Dominating Set} problem is \np-hard \cite{Bib:GAJO79}, and
the same holds for PDS \cite{Bib:GUNR05}. Liao and Lee
\cite{Bib:LILE05} proved that PDS on \emph{split} graphs is
\np-complete, and also they presented a polynomial time algorithm
for solving PDS optimally on \emph{interval} graphs. Dorfling and
Henning computed the power domination number, i.e.\ the size of
optimal power dominating set, for $n\times m$ grids
\cite{Bib:DOHE06}. Brueni and Heath \cite{Bib:BRHE05} have more
results on PDS, especially the \np-completeness of PDS on planar
bipartite graphs. To the best of our knowledge, no further results
are known on solving the PDS problem, either optimally or
approximately. Some of the results in this paper have appeared in
the thesis of the second author \cite{Bib:ST06}, and in the
proceedings of a workshop \cite{Bib:AAST07}.
\subsection{Our contributions}
Our results substantially improve on the understanding of PDS in the
context of approximation algorithms. In particular, we show a
substantial gap between the approximation guarantees for the
\textsc{Dominating Set} problem and PDS modulo a variant of the
$\p\not=\np$ conjecture. This seems to be the first known
``separation'' result between the two problems, in any class of
graphs.
\begin{itemize}
\item We present a reduction from the \textsc{MinRep} problem to the
PDS problem that  shows that PDS cannot be approximated within a
factor of $2^{log^{1-\epsilon}{n}}$, unless $\np\subseteq
\dtime(n^{polylog(n)})$.
\item For undirected graphs, we introduce the notion of strong regions and weak regions as a
means of obtaining lower bounds on the size of an optimal solution
for PDS. Based on this, we develop an approximation algorithm for
PDS that gives an approximation  guarantee of $O(k)$ for graphs that
have tree-width $k$. The algorithm requires the tree decomposition
as part of the input, and runs in time $O(n^3)$ (independent of
$k$). By slightly modifying this algorithm we get an algorithm that
solves PDS optimally  on trees. Our algorithm provides an
$O(\sqrt{n})$-approximation algorithm for PDS on planar graphs
because a tree decomposition of a planar graph with width
$O(\sqrt{n})$ can be computed efficiently \cite{Bib:ABFKN02}.
Moreover, we show that our methods (specifically, the lower bounds
used in our analysis) cannot improve on our $O(\sqrt{n})$
approximation guarantee.
\item We extend PDS in a natural way to directed graphs and
prove that even for {\it directed acyclic} graphs, PDS is hard to
approximate within the same threshold as for undirected graphs
modulo the same complexity assumption.
\item We give a linear-time algorithm based on dynamic-programming for
\textsc{Directed PDS} when the underlying undirected graph has
bounded tree-width. This builds on results and methods of Guo et
al.~ \cite{Bib:GUNR05}.
\end{itemize}
\section{PDS in Undirected Graphs\label{Sec:Hard}}
In this section we prove a result on the hardness of approximating
PDS by a reduction from the \textsc{MinRep} problem.  In Section
\ref{Sec:MinRep} we define the \textsc{MinRep} problem, and then we
give a \emph{gap preserving reduction} from \textsc{MinRep} to PDS
in Section \ref{Sec:UnDirHard}.

\subsection{The \textsc{MinRep} problem \label{Sec:MinRep}}
In the \textsc{MinRep} \cite{Bib:KOR01} problem we are given a
bipartite graph $G=(A,B,E)$ with a partition of $A$ and $B$ into
equal-sized subsets. Let $q_A$ and $q_B$ denote the number of sets
in the partition of $A$ and $B$, respectively. Let $A=A_1\cup
A_2\cup\cdots\cup A_{q_A}$ denote the partition of $A$, and let
$B=B_1\cup B_2\cup\cdots\cup B_{q_B}$ denote the partition of $B$.
This partition naturally defines a super bipartite graph
${\cal{H}}=({\cal{A}},{\cal{B}},{\cal{E}})$.
 The super nodes of $\cal{H}$ are ${A_1,A_2,\ldots,A_{q_A}}$ and
${B_1,B_2,\ldots,B_{q_B}}$. There is a super edge between super
nodes $A_i$ and $B_j$ if there exists some $a\in A_i$ and $b\in B_j$
such that $ab$ is an edge in $G$. We say that super edge $A_iB_j$ is
covered by nodes $a, b$ if $a\in A_i$, $b\in B_j$, and there is an
edge between $a$ and $b$ in $G$. Given $S\subseteq A\cup B$ we say
that the super edge $A_iB_j$ is covered by $S$ if there exists $a,
b\in S$ that covers $A_iB_j$. The goal in the \textsc{MinRep}
problem is to pick a minimum-size set of nodes, $A'\cup B'\subseteq
V(G)$, to cover all the super edges in ${\cal{H}}$. Note that we
need a pair of nodes to cover a super edge, and the pair should
induce an edge between the two super nodes of the super edge;
moreover, a node in $A'\cup B'$ may be useful for covering more than
one super edge. The following Theorem is from \cite{Bib:KOR01}.
\begin{theorem} [Theorem 5.4 in \cite{Bib:KOR01}]
The \textsc{MinRep} problem cannot be approximated within ratio
$2^{{log}^{1-\epsilon}{n}}$, for any fixed $\epsilon >0$, unless
$\np\subseteq \dtime(n^{polylog(n)})$.
\end{theorem}
\subsection{The reduction to PDS \label{Sec:UnDirHard}}
\begin{theorem}
\label{Thm:HardPDS} The PDS problem cannot be approximated within
ratio $2^{{log}^{1-\epsilon}{n}}$, for any fixed $\epsilon>0$,
unless $\np\subseteq \dtime(n^{polylog(n)})$.
\end{theorem}
\noindent{\bf The reduction:} Theorem \ref{Thm:HardPDS} is proved by
a reduction from the \textsc{MinRep} problem. We create an instance
$\overline{G}=(\overline{V},\overline{E})$ of the PDS problem from a
given instance $G=(A,B,E)(
{\cal{H}}=({\cal{A}},{\cal{B}},{\cal{E}})$) of the \textsc{MinRep}
problem. The idea is to replace each super edge with a  ``cover
testing gadget''.
\begin{enumerate}
\item Start with a copy of each node in $A\cup B$ in $\overline{G}$.
For convenience, we use the same notation for nodes (and set of
nodes) in $G$ and their copies in $\overline{G}$.
\item Add a new node $w^*$  to the graph $\overline{G}$, and connect
$w^*$ to all nodes in $A\cup B$. Also add new nodes $w^*_1, w^*_2,
w^*_3$ and connect them to $w^*$ (the nodes $w^*_1, w^*_2, w^*_3$
are added to force $w^*$ to be in any optimal solution. See the
proof of Lemma \ref{Lem:Red} for more details).
\item $\forall i\in\set{1,\ldots,q_A}, j\in\set{1,\ldots,q_B}$ if $A_iB_j$ is a super edge, then do
the following:
\begin{enumerate}
\item Let $E_{ij}$ denote the set of edges between $A_i$ and
$B_j$ in $G$ and let $\ell_{ij}$ denote $\card{E_{ij}}$ (see Figure
\ref{Fig:HardConsta}; for an example $E_{11}$ has $3$ edges, and
$E_{12}$ has $4$ edges). We denote the edges in $E_{ij}$ by $e_1,
e_2,\cdots,e_k,\cdots$.

\item Let $C_{ij}$ be a cycle of $3\ell_{ij}$ nodes. We sequentially
label the nodes of $C_{ij}$ as $u_1, v_1, w_1$, $u_2, v_2, w_2$,
$\cdots, u_k, v_k, w_k, \cdots$ (informally speaking, we associate
each triple $u_k, v_k, w_k$ with an edge $e_k$ of $E_{ij}$). Make
$\lambda=4$ new copies of the graph $C_{ij}$ ($\lambda$ can be any
constant greater than $3$; refer to the proof of Lemma \ref{Lem:Red}
for more details). For each edge $e_k=a_kb_k\in E_{ij}$ and for each
of the 4 copies of $C_{ij}$, we add an edge from $a_k$ to $u_k$ and
an edge from $b_k$ to $v_k$. See Figures \ref{Fig:HardPDS-1},
\ref{Fig:HardPDS-2} for an illustration.
\begin{figure}[!h]
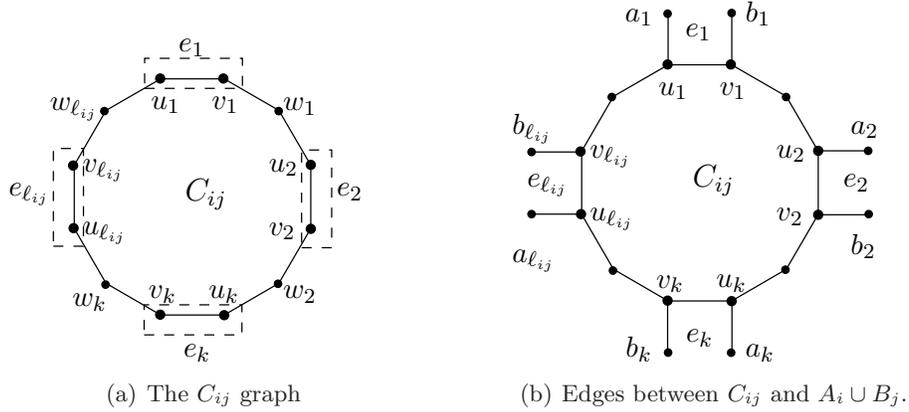

\begin{center}
\subfigure[The $C_{ij}$ graph] {
    \label{Fig:HardPDS-1}
    \input{HardPDS-1.pstex_t}
} \hspace{1cm} \subfigure[Edges between $C_{ij}$ and $A_i\cup B_j$.]
{
    \label{Fig:HardPDS-2}
    \input{HardPDS-2.pstex_t}
} \caption{The cover testing gadget.} \label{Fig:HardPDS}
\end{center}
\end{figure}
\end{enumerate}
\item Let $\overline{G}=(\overline{V},\overline{E})$ be the obtained
graph (see Figure \ref{Fig:HardConst} for an illustration).
\end{enumerate}
\begin{figure}[!hb]
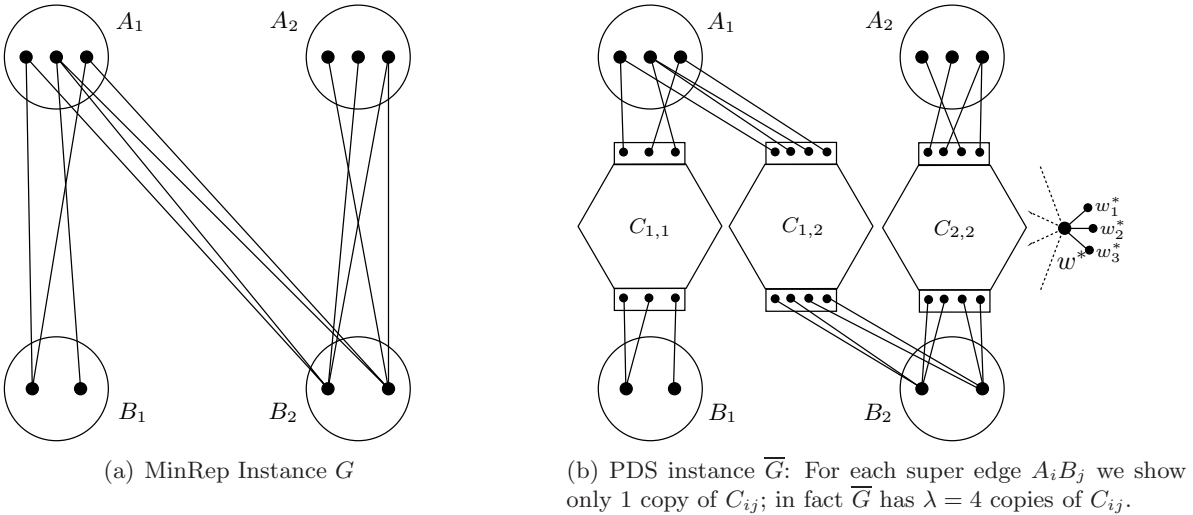

\begin{center}
\subfigure[MinRep Instance $G$] {\label{Fig:HardConsta}
    \input{HardConsta.pstex_t}
} \hspace{1.1cm} \subfigure[PDS instance $\overline{G}$: For each
super edge $A_iB_j$ we show only $1$ copy of $C_{ij}$; in fact
$\overline{G}$ has $\lambda=4$ copies of $C_{ij}$.] {
    \input{HardConstb.pstex_t}
} \caption{The hardness construction} \label{Fig:HardConst}
\end{center}
\end{figure}

Let $S$ be a feasible solution for the resulting PDS instance
$\overline{G}$, and suppose $w^*\in S$. Then all of the nodes in
$A\cup B$ are power dominated (by Rule 1 of PDS). Now consider a
gadget $C_{ij}$, and assume a node $v$ of $C_{ij}$ is in $S$. By
applying Rule 1 once and then repeatedly applying Rule 2 of PDS, the
gadget $C_{ij}$ will be completely power dominated, that is, all
nodes of the gadget will be in $\pd_S$.

The next lemma shows that the size of an optimal solution in PDS is
exactly one more than the size of an optimal solution in
\textsc{MinRep}. The number of nodes in the constructed graph is
equal to $\card{V(\overline{G})}=4+\card{V(G)}+3\lambda\card{E(G)}$.
This will complete the proof of Theorem \ref{Thm:HardPDS} by showing
that the above reduction is a gap preserving reduction from
\textsc{MinRep} to PDS with the same gap (hardness ratio) as the
\textsc{MinRep} problem.
\begin{lemma} \label{Lem:Red}
$A^*\cup B^*$ is an optimal solution to the instance $G=(A,B,E)$ of
the \textsc{MinRep} problem if and only if $S^*=A^*\cup
B^*\cup\set{w^*} \subseteq V(\overline{G})$ is an optimal solution
to the instance $\overline{G}$ of the PDS problem.
\end{lemma}
\begin{proof}
First, we claim that $w^*$ should be in any optimal solution of the
PDS instance $\overline{G}$. Suppose that $w^*$ is not in some
optimal solutions. Then, in order to power dominate the nodes $w^*,
w_1^*, w_2^*, w_3^*$ in $\overline{G}$, the set $S$ must contain at
least two of the nodes (leaves) $w_1^*, w_2^*, w_3^*$. This is a
contradiction, since we can replace these $2$ nodes by $w^*$ and
 obtain a smaller feasible solution.

Assume that $A^*\cup B^*$ is a feasible solution for the
\textsc{MinRep} instance $G$. We will show that $S=A^*\cup
B^*\cup\set{w^*}$ is a feasible solution to the PDS instance
$\overline{G}$. Note that all nodes in $A\cup B\cup \set{w^*, w^*_1,
w^*_2, w^*_3}$ are power dominated by applying Rule 1  on $w^*$.
Now, we only need to show that all nodes in the gadgets $C_{ij}$ are
power dominated. Consider any super edge $A_iB_j$ of ${\cal H}$. The
set $A^*\cup B^*$ covers all the super edges in ${\cal{H}}$. So
there exists a pair of nodes $a_k\in A^*\cap A_i$, $b_k\in B^*\cap
B_j$ that induces an edge of $G$. Since $a_k$ and $b_k$ are in $S$,
their neighbors, $u_k$ and $v_k$, in each of the $\lambda=4$ copies
of $C_{ij}$ in $\overline{G}$, will be power dominated by applying
Rule 1. Then the nodes $u_k$ and $v_k$ in each copy of $C_{ij}$ will
power dominate the entire cycle by repeatedly applying Rule 2. To
see this, note that any node in $C_{ij}$ has exactly $2$ neighbors
in $C_{ij}$ and at most $1$ neighbor not in $C_{ij}$. The neighbors
not in $C_{ij}$ are from $A_i\cup B_j$, and they are power dominated
by $w^*$. Hence, if a node in $C_{ij}$ and one of its neighbors in
$C_{ij}$ are power dominated, then by applying Rule 2 the other
neighbor in $C_{ij}$ will be power dominated. Hence, by starting
from $v_k$ and repeatedly applying Rule 2, we can sequentially power
dominate the nodes in $C_{ij}$. This shows that $S$ power dominates
all nodes in $\overline{G}$. Therefore, $\Opt(\overline{G})$ is at
most $\card{A^*\cup B^*}+1.$

Let $S^*\subseteq V(\overline{G})$ be an optimal solution for PDS.
By the above claim, $w^*$ is in $S^*$. Now define $A'=A\cap S^*$ and
$B'=B\cap S^*$. First we prove that any optimal solution of PDS is
contained in $A\cup B\cup\set{w^*}$, and then we show that $A'\cup
B'$ covers all super edges of the \textsc{MinRep} instance $G$.
Suppose that $S^*$ contains some nodes not in $A\cup B\cup
\set{w^*}$. Hence, there are some gadgets that are not completely
power dominated by $S^*\cap(A\cup B\cup \set{w^*})$. Let $C_{ij}$ be
such a gadget. By symmetry each of the $\lambda=4$ copies of
$C_{ij}$ is not completely power dominated.
 Therefore, the optimal solution $S^*$ needs to have
at least $3$ nodes from the $4$ copies of $C_{ij}$. By removing
these $3$ nodes from $S^*$ and adding $a_k\in A_i$ and $b_k\in B_j$
to $S^*$ for some arbitrary edge ${a_kb_k}\in E_{ij}$, we can power
dominate all of the $4$ copies of $C_{ij}$. This contradicts  the
minimality of $S^*$, and proves that $S^*\subseteq A\cup
B\cup\set{w^*}$. To see that $A'\cup B'$ covers all super edges,
note the following: suppose no node from any copy of $C_{ij}$ is in
the optimal solution; then any $C_{ij}$ can be power dominated only
by taking a pair of nodes $a\in A_i$, $b\in B_j$ that induces an
edge of $G$. This completes the proof of the lemma.
\end{proof}
\section{Approximation Algorithms for Planar
Graphs\label{Sec:Planar}} In this section we describe an
$O(k)$-approximation algorithm for PDS in graphs with tree-width
$k$; the running time is $O(n^3)$, independent of $k$. This
algorithm gives an $O(\sqrt{n})$-approximation algorithm for PDS in
planar graphs, since the tree-width of a planar graph $G$ with $n$
nodes is $O(\sqrt{n})$ and in $O(n^{\frac{3}{2}})$ time we can find
an $O(\sqrt{n})$ tree-width decomposition of the given planar graph
$G$ \cite{Bib:ABFKN02}. Finally, we show that the analysis of our
algorithm is tight on planar graphs. We use \textsc{Planar PDS} to
denote the special case of the PDS problem where the graph is
planar.

\begin{definition}\emph{\cite{Bib:DI00}}\label{Def:NTWD} A \emph{tree decomposition} of a graph
$G=(V,E)$ is a pair $\langle \set{X_i\subseteq V| i\in
I},T=(I,F)\rangle$ such that $T$ is a tree with $V(T)=I$, $E(T)=F$,
and satisfying the following properties:
\begin{enumerate}
\item[(T1)] $\bigcup_{i\in I}{X_i}=V$, and every edge ${uv}\in E$ has both ends in some $X_i$,
\item[(T2)] For all $i,j,k\in I$ if $j$ is on the unique path from $i$
to $k$ in $T$ then we have: $X_i\cap X_k\subseteq X_j$,
\end{enumerate}
The \emph{width} of $\langle \set{X_i| i\in I},T\rangle$ is the
$max_{i\in I}{\card{X_i}}-1$. The \emph{tree-width} of $G$ is
defined as the minimum width over all tree decompositions. The nodes
of the tree are called \emph{$T$-nodes} and the sets $X_i$ are
called \emph{bags}.
\end{definition}
{ A \emph{nice} tree decomposition is a tree decomposition $\langle
\set{X_i\subseteq V| i\in I},T=(I,F)\rangle$, where $T$ is a rooted
tree in which each node has at most $2$ children. If a node $i\in I$
has two children $j,k$ then $X_i=X_j=X_k$ ($i$ is called a
\textsc{Join} node), and if $i$ has one child $j$ then either
$X_j\subset X_i$ and $\card{X_i\setminus X_j}=1$ or $X_i\subset X_j$
and $\card{X_j\setminus X_i}=1$ ($i$ is called an \textsc{Insert} or
a \textsc{Forget} node, respectively).}

We introduce the notion of a strong region before presenting our
algorithm. Informally speaking, a set of nodes $R\subseteq V$ is
called strong if every feasible solution to the PDS problem has a
node of $R$. For a graph $G=(V,E)$, the neighborhood of $R\subseteq
V$ is $nbr(R)=\set{v\in V| \exists uv\in E,\ u\in R,\ v\notin R}$,
and the exterior of $R$ is defined by $ext(R)=nbr(V\setminus R)$,
i.e., $ext(R)$ consists of the nodes in $R$ that are adjacent to a
node in $V\setminus R$.
\begin{definition} Given a graph $G=(V,E)$ and a set $S\subseteq V$, the
subset $R\subseteq V$ is called an \emph{$S$-strong region} if
$R\not\subseteq \pd_{S\cup nbr(R)}$, otherwise, the set $R$ is
called an \emph{$S$-weak region}. The region $R$ is called
\emph{minimal $S$-strong} if it is an $S$-strong region and $\forall
r\in R$, $R-r$ is an $S$-weak region.
\end{definition}
It is easy to check from the definition that an $S$-strong region is
also an $\emptyset$-strong (or shortly strong) region. Any feasible
solution to the PDS problem needs to have at least one node from
every strong region.
\begin{lemma} \label{Lem:Alt}
A subset $R\subseteq V$ is an $S$-strong region if and only if for
every feasible solution $S\cup S^*$ of $G$,  we have $R\cap
(S^*\setminus S)\neq\emptyset$.
\end{lemma}
\begin{proof}
It can be seen that the set $S\cup(V\setminus R)$ will power
dominate the same set of nodes in $R$ that can be power dominated by
$S\cup nbr(R)$; this is valid for any subset $R\subseteq V$.

Let $R$ be an $S$-strong region. By the definition of a strong
region we have $R\not\subseteq \pd_{S\cup nbr(R)}$. Hence, by the
above claim $R\not\subseteq \pd_{S\cup (V\setminus R)}$. This shows
that every feasible solution $S^*$ needs to have at least one node
from $R$ that is not in $S$.

Now assume that for every feasible solution $S^*$ of $G$ we have
$R\cap(S^*\setminus S)\neq\emptyset$. Suppose that $R$ is an
$S$-weak region, so by the definition of a weak region we have
$R\subseteq \pd_{S\cup nbr(R)}$. It follows that
$S^*=S\cup(V\setminus R)$ is a feasible solution, but $R$ has no
intersection with $S^*\setminus S\subseteq (V\setminus R)$. This is
a contradiction, so $R$ is an $S$-strong region.
\end{proof}

Our algorithm makes one level-by-level and bottom-to-top pass over
the tree $T$ of the tree decomposition of $G$ and constructs a
solution $S$ for PDS (initially, $S=\emptyset$). At each node $r_j$
of $T$ we check whether the union of the bags in the subtree rooted
at $r_j$ forms an $S$-strong region; if yes, then the bag $X_{r_j}$
of $r_j$ is added to $S$, otherwise $S$ is not updated. The key
point in the analysis is to show that $\Opt(G)\geq m$, where $m$ is
the number of nodes of $T$ where we updated $S$.
\begin{algorithm}[h]
\caption{$O(k)$-approximation Algorithm}
 \label{Alg:TWD}
\begin{algorithmic}[1]
  \STATE A tree decomposition $\langle \set{X_i| i\in
  I},T\rangle$ of $G$ is given, where $T$ is rooted at $r$.
  \STATE Let $I_{\ell}$ be the set of $T$-nodes at distance $\ell$
  from the root, and let $d$ be the maximum distance from $r$ in $T$.
  \STATE $S\leftarrow\emptyset$
  \FOR{$i=d$ to $0$}
    \STATE Let $I_i=\{r_1,\ldots,r_{k_i}\}$ and denote by $T_{r_j}$ the subtree in $T$ rooted at
    $r_j$.
    \STATE Let $Y_{r_j}$ be the union of bags corresponding to the $T$-nodes in
    $T_{r_j}$.
    \FOR{$j=1$ to $k_i$}
        \IF{ $Y_{r_j}$ is an $S$-strong region}
            \STATE $S\leftarrow S\cup X_{r_j}$; where $X_{r_j}$ is the bag corresponding to $r_j$.
%            \STATE $a\leftarrow a+1$
 %           \STATE $\ST_a=Y_j\setminus\bigcup_{s=1}^{a-1}\ST_s$
        \ENDIF
    \ENDFOR
  \ENDFOR
  \STATE Output $S_o=S$
\end{algorithmic}
\end{algorithm}

\subsection{Analysis of the algorithm}
In this subsection we show that our algorithm has an approximation
guarantee of $O(k)$. Let $G=(V,E)$ denote the input graph, and let
$S\subseteq V$ be any set of nodes.

\begin{lemma}\label{Lem:WKRGN}
Suppose $Z$ is an $S$-weak region such that $ext(Z)\subseteq S$.
Then we have $Z\subseteq \pd_{S}.$
\end{lemma}
\begin{proof}
Let $Y=ext(Z)$, it is easy to check that $nbr(Z\setminus Y)\subseteq
ext(Z)$. We claim that $Z\setminus Y$ is an $S$-weak region. Let
$S^*=V\setminus (Z\cup S)$, it is easy to check that $S\cup S^*$ is
a feasible solution for the graph $G$, but $S^*\cap (Z\setminus
Y)=\emptyset$. Hence, by Lemma \ref{Lem:Alt}, $Z\setminus Y$ is not
an $S$-strong region, and so it is an $S$-weak region. Thus
$Z\setminus Y\subseteq \pd_{S\cup nbr(Z\setminus Y)}\subseteq
\pd_{S\cup ext(Z)}=\pd_{S}$ and this implies that $Z\subseteq
\pd_{S}$ as $Y=ext(Z)\subseteq S$.

\end{proof}
\begin{lemma}\label{Lem:StrngReg}
Let $Z\subseteq V$ be an $S$-strong region. Suppose that $Y$ is a
subset of $V$ such that $Y\subseteq\pd_{S}$ and $ext(Y)\subseteq S$.
Then $Z\setminus Y$ is an $S$-strong region.
\end{lemma}
\begin{proof}
Assume for the sake of contradiction that $Z\setminus Y$ is an
$S$-weak region. Then by the definition of strong regions we have: $
Z\setminus Y\subseteq \pd_{S\cup nbr(Z\setminus Y)}$. It is easy to
see that $nbr(Z\setminus Y)\subseteq nbr(Z)\cup ext(Y)$. This
implies that $ Z\setminus Y\subseteq \pd_{S\cup nbr(Z\setminus
Y)}\subseteq \pd_{S\cup nbr(Z)\cup ext(Y)}=\pd_{S\cup nbr(Z)}$. The
condition in the lemma states that $Y\subseteq \pd_{S}\subseteq
\pd_{S\cup nbr(Z)}$. Hence, we get $Z=(Z\setminus Y)\cup (Z\cap
Y)\subseteq \pd_{S\cup nbr(Z)}$, which means that $Z$ is an $S$-weak
region. This is a contradiction, so the lemma is proved.
\end{proof}

\begin{theorem}\label{Thm:TWDKALG}
Given a graph $G=(V,E)$ and a tree decomposition of $G$ of width $k$
as input, Algorithm \ref{Alg:TWD} runs in time $O(n\cdot \card{E})$,
and achieves an approximation guarantee of $(k+1)$.
\end{theorem}
\begin{proof}
First, we show that the solution $S_o$ found by the algorithm is
feasible. Then we prove the approximation guarantee, and establish
the running time.

For any node $q$ of $T$, recall that $Y_q$ denotes the union of the
bags corresponding to the $T$-nodes in the subtree rooted at $q$ in
$T$; let $G_q$ denote the subgraph of $G$ induced by $Y_q$. We claim
that $ext(Y_q)\subseteq X_q$. Suppose that $q$ has $m$ children in
$T$, call them $c_1, \ldots, c_m$. For each edge $qc_{j}$ ($j=1,
\cdots, m$), the set $X_q\cap X_{c_j}$ {\it separates} $Y_{c_j}$
from the rest of the graph, that is, every path between a node in
$Y_{c_j}$ and a node in $V\setminus Y_{c_j}$ contains a node of
$X_q\cap X_{c_j}$ (see Lemma 12.3.1 in \cite{Bib:DI00}). Thus,
$ext(Y_{c_j})\subseteq X_q\cap X_{c_j}\subseteq X_q$, and hence, for
$Y_q=X_q\cup Y_{c_1}\cup\cdots\cup Y_{c_m}$, we have
$ext(Y_q)\subseteq X_q$.

We use induction on the height of the subtree of $T$ rooted at $q$
to prove the following: if $Y_q$ is $S^*$-strong, then
$Y_q\subseteq\pd_{S^*\cup X_q}$, where $S^*$ denotes the solution
just before the algorithm examines $Y_q$. The statement clearly
holds when $q$ is a leaf of $T$ (since $Y_q=X_q$). Otherwise, let
$c_1, \ldots, c_m$ be the children of $q$ in $T$. For each $j=1,
\ldots, m$, when the algorithm examined $Y_{c_j}$, either $Y_{c_j}$
was $S$-weak, in which case (by Lemma \ref{Lem:WKRGN}) we have
$Y_{c_j}\subseteq \pd_{S\cup ext(Y_{c_j})} \subseteq
\pd_{S\cup(X_{c_j}\cap X_q)}\subseteq \pd_{S^*\cup X_q}$ or
$Y_{c_j}$ was $S$-strong in which case $Y_{c_j}\subseteq \pd_{S\cup
X_{c_j}}$ by induction (note that $S\cup X_{c_j}\subseteq S^*$); we
use $S$ to denote the solution just before the algorithm examines
$Y_{c_j}$. Hence, $Y_q=Y_{c_1}\cup \cdots \cup Y_{c_m}\cup
X_{q}\subseteq \pd_{S^*\cup X_q}$.

The above statement implies that $V\subseteq \pd_{S_o}$ because at
the step when the algorithm examines the root $r$ of $T$ either
\begin{itemize}
\item[(i)] $Y_r$ is $S$-strong, so $S_o=S\cup X_r$, and
$Y_r\subseteq \pd_{S\cup X_r}=\pd_{S_o}$; or
\item[(ii)] $Y_r$ is $S$-weak, and $Y_r\subseteq \pd_{S\cup
ext(Y_r)}=\pd_{S_o}$; since $Y_r=V(G)$ and $ext(Y_r)=\emptyset$.
\end{itemize}

To show that the approximation guarantee is $(k+1)$ we will
construct a set $\R$ of pairwise disjoint strong regions $R_1, R_2,
\ldots,$ such that there is a strong region $R_j$ corresponding to
each step of the algorithm that adds a non empty bag $X_{q_j}$ to
$S$. Thus $\card{S_o}\leq (k+1)\card{{\R}}$ since each bag has $\leq
k+1$ nodes, and $\Opt(G)\geq\card{{\R}}$ because  every feasible
solution has size $\geq\card{\R}$, by Lemma \ref{Lem:Alt}. Hence,
$\card{S_o}\leq (k+1)\Opt(G)$. We construct the sets $R_1, R_2,
\ldots,$ during the execution of the algorithm as follows. Suppose
the algorithm finds $Y_q$ to be $S$-strong while examining a node
$q$ of $T$. Let $q_1,\ldots, q_{\ell-1}$ be the nodes of $T$ where
the algorithm updated the solution before examining $q$, and let $S$
be the solution just before the algorithm examines $q$. Then define
$R_{\ell}=Y_{q_{\ell}}\setminus(Y_{q_1}\cup\cdots\cup
Y_{q_{\ell-1}})$, where ${q_\ell}=q$.  We claim that $R_{\ell}$ is
an $S$-strong region.  For each strong region $Y_{q_j}$ ($j=1,
\ldots, \ell-1$) we have seen that $ext(Y_{q_j})\subseteq
X_{q_j}\subseteq S$ and $Y_{q_j}\subseteq \pd_S$; note that the
algorithm added $X_{q_j}$ to the solution since $Y_{q_j}$ was a
strong region. It follows  that $ext(Y_{q_1}\cup\cdots\cup
Y_{q_{\ell-1}})\subseteq S$, and $Y_{q_1}\cup\cdots\cup
Y_{q_{\ell-1}}\subseteq \pd_S$. Hence,  by Lemma \ref{Lem:StrngReg},
the set $R_{\ell}$ is an $S$-strong region. Clearly, the sets $R_1,
R_2,\ldots,$ are pairwise disjoint. This completes the construction
of $\R$.

Consider the running time. Without loss of generality we can assume
that the given tree decomposition of width $k$ has at most $4n$ bags
(see Lemma 13.1.2 in \cite{Bib:KL94}). Using standard algorithmic
techniques   we can test in $O(\card{E})$ time whether a given set
$R\subseteq V$ is an $S$-strong region (we compute $\pd_{S\cup
nbr(R)}$ and check if it contains $R$). Therefore, our algorithm has
a running time of $O(n\cdot \card{E})$.
\end{proof}

It is known that planar graphs have tree-width $O(\sqrt{n})$, and
such a tree decomposition can be found in $O(n^{\frac{3}{2}})$ time
\cite{Bib:ABFKN02}. This fact together with the above theorem proves
the following theorem.
\begin{theorem}
Algorithm \ref{Alg:TWD} achieves an approximation guarantee of
$O(\sqrt{n})$ for the \textsc{Planar PDS} problem.
\end{theorem}

As mentioned earlier, Haynes et al.\cite{Bib:HAHHH02} presented a
linear-time algorithm for optimally solving PDS on trees. By
modifying Algorithm \ref{Alg:TWD}, we can solve PDS optimally on
trees. The resulting algorithm differs from the algorithm of Haynes
et al.\ since our algorithm uses strong regions. Informally, the
algorithm makes a level-by-level and bottom-to-top pass over the
tree $G$. At a node $v$ of the tree $G$ if the set of nodes in the
subtree rooted at $v$ forms a strong region, then we add $v$ to the
solution, otherwise we skip $v$. Formally, we define $X_v=\set{v}$
for each $v\in V(G)$, and we run Algorithm \ref{Alg:TWD} on the tree
$G$. Note that defining bags in this way does not give a tree
decomposition of $G$.

\begin{theorem}
A modification of Algorithm \ref{Alg:TWD} runs in time
$O(n\cdot\card{E})$ and solves PDS optimally on trees.
\end{theorem}
\subsection{\bf Lower bounds via disjoint strong regions} In this part we show
that any approximation algorithm for PDS that uses the number of
disjoint strong regions as a lower bound has an approximation
guarantee of $\Omega(\sqrt{n})$. In proposition \ref{Pro:GridOPT} we
give a lower bound on the optimal value for PDS on an $\ell\times m$
grid. Independently, \cite{Bib:DOHE06} gave a stronger result for
this.

\begin{lemma} Any minimal $S$-strong region is connected.
\label{Lem:MinContRgn}
\end{lemma}
\begin{proof}
Assume that $R$ is a minimal $S$-strong region that is not
connected. Let $C\subset R$ be a connected component of R. The set
$C$ is an $S$-weak region since $R$ is a minimal $S$-strong region.
By the definition of a weak region we have $C\subseteq \pd_{S\cup
nbr(C)}$. The set $C$ is a connected component of $R$, so the
neighborhood of $C$ has no intersection with $R\setminus C$. This
implies that $nbr(C)\subseteq nbr(R)$, and consequently we have
$C\subseteq \pd_{S\cup nbr(C)}\subseteq \pd_{S\cup nbr(R)}$. The
same argument as above shows that $R\setminus C\subseteq \pd_{S\cup
nbr(R)}$. Hence, $R \subseteq \pd_{S\cup nbr(R)}$ which is a
contradiction.
\end{proof}
\begin{lemma}
\label{Lem:MaxGridDSTR} The number of disjoint strong regions in an
$\ell\times m$ grid is exactly one.
\end{lemma}
\begin{proof}
For the sake of contradiction, assume that the given grid has two
disjoint strong regions. Take as few nodes as possible from these
strong regions until we get minimal strong regions, say $R_1$ and
$R_2$. It is easy to check that the set of nodes of any row or any
column of the grid power dominates all nodes in the grid. By Lemma
\ref{Lem:Alt}, $R_1$ and $R_2$ should have at least one node from
every feasible solution. In the other words, $R_1$ and $R_2$ must
have at least one node from each row and also from each column. By
lemma \ref{Lem:MinContRgn}, we know that $R_1$ and $R_2$ induce
connected subgraphs. Hence, in $R_1$ there is a path from a node in
the top row to a node in the bottom row, and also in $R_2$ there is
a path from a node in the rightmost column to a node in the leftmost
column. Obviously these two paths share a common node. This is a
contradiction, since $R_1$ and $R_2$ are assumed to be disjoint.
\end{proof}

We denote by $\pd^{i}_{S}$ a set of nodes that are power dominated
after applying propagation rule, Rule 2, for $i$ number of times to
$\pd^{0}_{S}=S\cup nbr(S)$. Obviously this depends on the order of
applying the propagation rule. We use the notation without
specifying the order of applying the propagation rule.
\begin{lemma}\emph{(Propagation
lemma)} Given an ordering of propagation rules applied to $S\cup
nbr(S)$ with $\pd_{S}^k=\pd_S$ we have:
$\card{ext(\pd_{S}^j)}\leq\card{ext(\pd_{S}^i)}, \forall\ 0\leq i <
j\leq k$. \label{Lem:PrpgLem}
\end{lemma}
\begin{proof}
We will prove that
$\left|ext(\pd_{S}^{i+1})\right|\leq\left|ext(\pd_{S}^i)\right|$,
for all $0\leq i\leq k-1$. Consider the set $\pd_{S}^i$ and assume
that in the $(i+1)$-{st} step we apply Rule 2 to $v\in
ext(\pd_{S}^i)$ and power dominate $u$; i.e. $u\in
\pd_{S}^{i+1},u\notin \pd_{S}^i$. To apply Rule 2 to $v$ all
neighbors of $v$ except $u$ should be power dominated, so we have
$nbr(v)\setminus \pd_{S}^i=\{u\}$. Also since we power dominate $u$
at step $(i+1)$, we have $nbr(v)\subseteq \pd_{S}^{i+1}$. Therefore,
$v$ is not in $ext(\pd_{S}^{i+1})$, but $u$ may be in the exterior
of $\pd_{S}^{i+1}$. Hence, we have
$ext(\pd_{S}^{i+1})\subseteq\left(ext(\pd_{S}^i)\setminus\set{v}\right)\cup\{u\}$.
It follows that $\left|ext(\pd_{S}^{i+1})\right| \leq
\card{ext(\pd_{S}^i)\setminus\set{v}}+\card{\set{u}}
=\left|ext(\pd_{S}^i)\right|$.
\end{proof}
\begin{proposition}\label{Pro:GridOPT}
Let $G$ be an $\ell\times m$ grid with $\ell\leq m$, then
$\Opt(G)=\Theta(\ell)$.
\end{proposition}
\begin{proof}
First note that any row or any column of the grid power dominates
all nodes. In the following we prove that any feasible solution of
PDS needs to have at least $\frac{l-1}{5}$ nodes.
 Assume  that
there exists $S\subseteq V$ such that $\card{S}<\frac{l-1}{5}$ and
$\pd_S=V(G)$. The maximum degree in $G$ is $4$, so we have
$\card{ext(\pd_{S}^0)}\leq \card{S \cup nbr(S)}<\frac{l-1}{5}\cdot
5=l-1$. Therefore, $\pd_{S}^0$ contains no full row or no full
column of $G$. The set $S$ power dominates $G$, so there is an $i$
such that $\pd_{S}^i$ contains a full row or a full column. Consider
the smallest $i$ with this property. Hence, $\pd_{S}^{i-1}$ has no
full row or full column. But some row or some column must have at
least $l-1$ nodes in $\pd_{S}^{i-1}$ since $\pd_{S}^i$ contains an
entire row or an entire column. Without loss of generality assume
that these $l-1$ nodes are from a column. By the definition of $i$,
each of the $l-1$ nodes in this column are in a row which is not a
subset of $\pd_{S}^{i-1}$. Therefore, there are at least $l-1$ rows
with at least one node in $\pd_{S}^{i-1}$ and at least one node not
in $\pd_{S}^{i-1}$. This implies that $\card{ext(\pd_{S}^{i-1})}\geq
l-1$. Finally, by using Lemma \ref{Lem:PrpgLem} we get the following
contradiction:
$l-1>\left|\pd_{S}^0\right|\geq\left|ext(\pd_{S}^0)\right|
\geq\left|ext(\pd_{S}^{i-1})\right|\geq l-1$.
\end{proof}

Consider any approximation algorithm for PDS that uses only the
number of disjoint strong regions as a lower bound on the size of an
optimal solution. By Lemma \ref{Lem:MaxGridDSTR}, this algorithm
finds a lower bound of $1$ on the size of an optimal solution on a
grid. The $\sqrt{n}\times\sqrt{n}$ grid has an optimal solution of
size $\Theta(\sqrt{n})$ by Proposition \ref{Pro:GridOPT}. This shows
that the approximation guarantee of the algorithm is
$\Omega(\sqrt{n})$, even on planar graphs.

\begin{proposition} \label{Pro:LBAP} Consider any approximation
algorithm for PDS that uses \emph{only} the number of disjoint
strong regions as a lower bound on the optimal value. Then the
approximation guarantee is $\Omega(\sqrt{n})$.
\end{proposition}
\section{PDS in Directed Graphs\label{Sec:Directed}}
In this section we extend the PDS problem to directed graphs to
obtain  the \textsc{Directed Power Dominating Set} (\textsc{Directed
PDS}) problem. Our motivation for studying the directed problem
comes from theoretical considerations. The \textsc{Dominating Set}
problem is studied on both undirected and directed graphs, and there
is extensive literature on the latter (see
\cite{Bib:HAHESL98a,Bib:HAHESL98b}). The similarities between the
\textsc{Dominating Set} problem and the PDS problem led us to define
and study the \textsc{Directed PDS} problem. We give a result on the
hardness of approximation of \textsc{Directed PDS}. Then we
reformulate the \textsc{Directed PDS} problem in terms of valid
coloring of the edges. Using this, we design an algorithm for
solving \textsc{Directed PDS} in linear-time on a special class of
directed graphs.

Let $G=(V,E)$ be a directed graph. A node $w$ is called an
out-neighbor (in-neighbor) of a node $v$ if there is a directed edge
from $v$ to $w$ (from $w$ to $v$) in $G$.  The number of
out-neighbors (in-neighbors) of a node $v$ is called the out-degree
(in-degree) of $v$ and is denoted by $d^+_G(v)$ (or similarly
$d^-_G(v)$). For a set of nodes $X$, the subgraph of $G$ induced by
$X$ is denoted by $G[X]$.
 The directed
graphs that we consider here have no loops nor parallel edges, but
may have two edges with different directions on the same two end
nodes (we call such edges \emph{antiparallel}). Given a directed
graph $G$ by the \emph{underlying undirected graph} we mean the
undirected graph obtained from $G$, by removing the direction of
edges and also removing any parallel edges that are introduced after
removing the directions.

\begin{definition}[the \textsc{Directed PDS} problem] Let $G$ be a directed graph. Given a set of nodes $S\subseteq V(G)$, the set of nodes that are power
dominated by $S$, denoted by $\pd_{S}$, is obtained as follows:
\begin{enumerate}
\item[(D1)] if node $v$ is in $S$, then $v$ and all of its
out-neighbors are in $\pd_{S}$;
\item[(D2)] (propagation) if node $v$ is in $\pd_{S}$, one of its out-neighbors $w$ is
not in $\pd_{S}$, and all other out-neighbors of $v$ are in
$\pd_{S}$, then $w$ is inserted into $\pd_{S}$.
\end{enumerate}
We say that $S$ power dominates $G$ if $\pd_{S}=V(G)$. The
\textsc{Directed PDS} problem is to find a node set $S$ with minimum
size that power dominates all the nodes in $G$.
\end{definition}
We prove a threshold of $2^{\log{n}^{1-\epsilon}}$ for the hardness
of approximation of \textsc{Directed PDS} modulo the same complexity
assumption as in Theorem \ref{Thm:HardPDS}. The proof uses a
reduction from the \textsc{MinRep} problem to the \textsc{Directed
PDS} problem in a  directed \emph{ acyclic} graph. This reduction is
similar to the reduction in Theorem \ref{Thm:HardPDS}; the main
difference comes from the gadget for modeling the super edges.
\begin{theorem} \label{Thm:HardDPDS}
The \textsc{Directed PDS} problem even when restricted to directed
\emph{ acyclic} graphs cannot be approximated within ratio
$2^{\log^{1-\epsilon}{n}}$, for any fixed $\epsilon > 0$, unless
$NP\subseteq DTIME(n^{polylog(n)})$.
\end{theorem}
\noindent{\bf The reduction:} We create an instance
$\overline{G}=(\overline{V},\overline{E})$ of the \textsc{Directed
PDS} problem from a given instance $G=(A,B,E)
({\cal{H}}=({\cal{A}},{\cal{B}},{\cal{E}})$)  of the \textsc{MinRep}
problem.

\begin{enumerate}
\item Start with a copy of each node in $A\cup B$ in $\overline{G}$.
For convenience, we use the same notation for nodes (and sets of
nodes) in $G$ and their copies in $\overline{G}$.
\item Add a new node $w^*$ to the graph $\overline{G}$, and add a
directed edge from $w^*$ to each node in $A\cup B$.
\item $\forall i\in\set{1,\ldots,q_A}, j\in\set{1,\ldots,q_B}$ if $A_iB_j$ is a super edge, then do
the following:
\begin{enumerate}
\item Let $E_{ij}$ be the set of edges between $A_i$ and $B_j$ in
$G$, and let $\ell_{ij}$ denote $\card{E_{ij}}$. We denote the edges
in $E_{ij}$ by $e_1,e_2,\ldots,e_k,\ldots,e_{\ell_{ij}}$.
\item  Let $D_{ij}$ be the graph on $6\ell_{ij}+1$ nodes as shown in Figure \ref{Fig:HardDPDS}.
In $D_{ij}$ there are  $6$ nodes $u_k, v_k, d_k, \alpha_k, \beta_k,
\gamma_k$ associated with an edge $e_k$ of $E_{ij}$. The part of
$D_{ij}$ associated with an edge $e_k$ is shown in Figure
\ref{Fig:HardDPDS-part}; note that all these parts share a common
node, called the \emph{center node}, in $D_{ij}$. Make $\lambda=4$
new copies of the graph $D_{ij}$ ($\lambda$ can be any constant
greater than $3$). For each edge $e_k=a_kb_k\in E_{ij}$ and for each
of the $4$ copies of $D_{ij}$, we add a directed edge from $a_k$ to
$u_k$ and a directed edge from $b_k$ to $v_k$. In addition to these
edges, there are directed edges from $w^*$ to some nodes inside
$D_{ij}$; these directed edges are denoted by a dashed line in
Figure \ref{Fig:HardDPDS}.

\begin{figure}[!h]
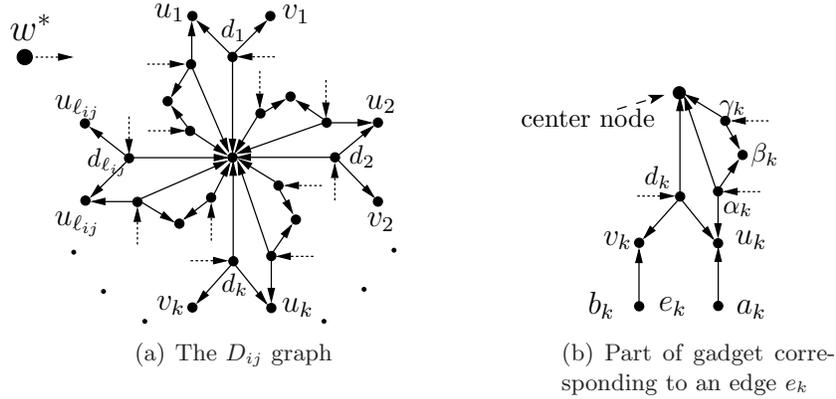

\begin{center}
\subfigure[The $D_{ij}$ graph] {
    \label{Fig:HardDPDS}
    \input{HardDPDS-1.pstex_t}
} \hspace{1cm} \subfigure[Part of gadget corresponding to an edge
$e_k$] {
    \label{Fig:HardDPDS-part}
    \input{HardDPDS-part.pstex_t}
} \caption{The cover testing gadget.}
\end{center}
\end{figure}
\end{enumerate}
\item Let $\overline{G}=(\overline{V},\overline{E})$ be the obtained
graph.
\end{enumerate}
The next lemma shows that the size of an optimal solution in
\textsc{Directed PDS} is exactly one more than the size of an
optimal solution in the \textsc{MinRep} instance. The number of
nodes in the constructed graph is at most
$\card{V(\overline{G})}\leq 1+\card{V(G)}+7\lambda\card{E(G)}$. This
will complete the proof of Theorem \ref{Thm:HardDPDS} by showing
that the above reduction is a \emph{gap preserving} reduction from
\textsc{MinRep} to \textsc{Directed PDS} with the same gap (hardness
ratio) as the \textsc{MinRep} problem.
\begin{lemma} \label{Lem:DRed}
$A^*\cup B^*$ is an optimal solution to the instance $G=(A,B,E)$ of
the \textsc{MinRep} problem if and only if $S^*=A^*\cup
B^*\cup\set{w^*} \subseteq V(\overline{G})$ is an optimal solution
to the instance $\overline{G}$ of the \textsc{Directed PDS} problem.
\end{lemma}
\begin{proof}
First note that  $w^*$ should be in any feasible solution; because
it has in-degree zero.

Assume that $A^*\cup B^*$ is a feasible solution for the
\textsc{MinRep} instance $G$. We will show that $S=A^*\cup
B^*\cup\set{w^*}$ is a feasible solution to the \textsc{Directed
PDS} instance $\overline{G}$. Note that all nodes in $A\cup B$ and
some nodes inside the gadgets $D_{ij}$ are power dominated by
applying rule (D1) on $w^*$. Now, we only need to show that all
nodes in the gadgets $D_{ij}$ are power dominated. Consider a super
edge $A_iB_j$ of  ${\cal H}$. The set $A^*\cup B^*$ covers all the
super edges in ${\cal{H}}$. Hence, there exists a pair of nodes
$a_k\in A^*\cap A_i$, $b_k\in B^*\cap B_j$ that induces an edge of
$G$. Since $a_k$ and $b_k$ are in $S$ their out-neighbors $u_k$ and
$v_k$ in each of the  $\lambda=4$ copies of the $D_{ij}$ graph will
be power dominated by applying rule (D1). Now node $d_k$, that is
already power dominated by $w^*$, will power dominate the center
node by an application of rule (D2). Now we claim that the center
node will power dominate the remaining nodes in the gadget $D_{ij}$.
Consider the part of gadget (shown in Figure
\ref{Fig:HardDPDS-part}) corresponding to an edge $e_r\in E_{ij}$.
Note that the nodes $\gamma_r$, $\alpha_r$, and $d_r$ are already
power dominated by $w^*$. It is easy to check that the nodes
$\beta_r, u_r, v_r$ will be power dominated by sequentially applying
rule (D2) on $\gamma_r$, $\alpha_r$, and $d_r$. This shows that $S$
power dominates all nodes in $\overline{G}$. Therefore,
$\Opt(\overline{G})$ is at most $\card{A^*\cup B^*}+1.$

Let $S^*\subseteq V(\overline{G})$ be an optimal solution for
\textsc{Directed PDS}. As we showed above $w^*$ should be in any
feasible solution for \textsc{Directed PDS}. Now define $A'=A\cap
S^*$ and $B'=B\cap S^*$. First we prove that any optimal solution of
\textsc{Directed PDS} is contained in $A\cup B\cup\set{w^*}$, and
then we show that $A'\cup B'$ covers all the super edges of ${\cal
H}$. Suppose that $S^*$ contains some nodes not in $A\cup B\cup
\set{w^*}$. Hence, there are some gadgets that are not completely
power dominated by $S^*\cap(A\cup B\cup \set{w^*})$. Let $D_{ij}$ be
such a gadget. By symmetry each of the $\lambda=4$ copies of
$D_{ij}$ is not completely power dominated.  Therefore, the optimal
solution $S^*$ needs to have at least $3$ nodes from the $4$ copies
of $D_{ij}$.
 By removing these $3$ nodes from $S^*$ and adding $a_k\in A_i$ and
$b_k\in B_j$ to $S^*$ for some arbitrary edge $e_k={a_kb_k}\in
E_{ij}$, we can power dominate all $4$ copies of $D_{ij}$. This
contradicts the optimality of $S^*$, and proves that $S^*\subseteq
A\cup B\cup\set{w^*}$. To see that $A'\cup B'$ covers all super
edges, it is enough to note the following. Suppose no node from any
copy of $D_{ij}$ is in the optimal solution, then any $D_{ij}$ can
be power dominated only by taking a pair of nodes $a\in A_i$ and
$b\in B_j$ that induces an edge of $G$. Otherwise,  any power
dominated node in the gadget has at least $2$ out-neighbors that are
not power dominated, so rule (D2) cannot be applied. This completes
the proof of the lemma.
\end{proof}

There are several notions for the tree-width of directed graphs such
as DAG width \cite{Bib:OB06}, directed tree-width
\cite{Bib:JORST01}, and Kelly-width \cite{Bib:HUKR07}. Directed
acyclic graphs have width equal to zero for the first two notions
\cite{Bib:OB06}, and have Kelly-width of $1$. Hence, Theorem
\ref{Thm:HardDPDS} gives a hardness threshold of
$O(2^{\log{n}^{1-\epsilon}})$ even if the directed graph has width
$\leq 1$ according to any of the above three notions.

We reformulate \textsc{Directed PDS} in terms of  \emph{valid
colorings} of the edges in order to develop an algorithm based on
dynamic-programming for \textsc{Directed PDS}. Guo et al.\
\cite{Bib:GUNR05} introduced the notion of valid orientations to get
a new formulation for PDS (in undirected graphs). They also designed
a linear-time dynamic-programming algorithm based on valid
orientations for optimally solving PDS on  graphs of bounded
tree-width. Our method applies to directed graphs such that the
underlying undirected graph has bounded tree-width.

\begin{definition} A \emph{coloring} of a directed graph $G=(V,E)$ is a partitioning of the edges in $G$  into
red and blue edges. We denote a coloring by $\mathcal{C}=(V,E_r\cup
E_b)$ where $E_r$ is the set of red edges and $E_b$ is the set of
blue edges.
\end{definition}\label{Def:ValidColor}
We reformulate the \textsc{Directed PDS} problem via a so-called
valid coloring of directed graphs; informally speaking, these
colorings ``model'' the application of rules (D1) and (D2) of
\textsc{Directed PDS}.
\begin{definition}
A \emph{valid coloring} $\mathcal{C}=(V,E_r\cup E_b)$ of a directed
graph $G=(V,E)$ is a coloring of $G$ with the following properties:
\begin{enumerate}
\item No two antiparallel edges can be colored red.
\item The subgraph induced by the red edges, $G_r=(V,E_r)$, has the following properties:
\begin{enumerate}
\item $\forall v\in G:\, d^{-}_{G_r}(v)\leq 1$, and
\item $\forall v\in G: \, d^{-}_{G_r}(v)=1 \Longrightarrow d^{+}_{G_r}(v)\leq 1$.
\end{enumerate}
\item $G$ has no \emph{dependency cycle}. A dependency cycle is a sequence of
directed edges whose underlying undirected graph forms a cycle such
that all  the red edges are in one direction, all the blue edges are
in the other direction,  and there are no two consecutive blue
edges.
\end{enumerate}
We call a node an \emph{origin} of $\mathcal{C}$ if it has no
incoming edges in $G_r$.
\end{definition}
Our dynamic-programming algorithm for \textsc{Directed PDS} is based
on the following lemma.
\begin{lemma}\label{Lem:ValidColor}
Given a directed graph $G$ and $S\subseteq V(G)$, $S$ power
dominates $G$ if and only if there is a valid coloring of $G$ with
$S$ as the set of origins.
\end{lemma}
\begin{proof}
Suppose $S\subseteq V$ power dominates $G$. Then we give a valid
coloring $\mathcal{C}$ with $S$ as the set of origins by coloring
the edges in $G$ according to the way that $S$ power dominates $G$.
We color an edge $(v,w)$ red if node $w$ is power dominated by
applying the power domination rules on $v$; either by the domination
rule (D1) or by the propagation rule (D2). Note that when we apply
the propagation rule (D2), then we do not power dominate the
previously power dominated nodes. Also when we apply rule (D1) on
$v$, then we power dominate all (not some subset of) neighbors of
$v$ that are not power dominated. We write $v < u$ when a node $u$
is power dominated after $v$. It is easy to check that with this
coloring the degree requirements are satisfied; each node can be
power dominated only once, and if it is a power dominated node (not
in $S$) , then it cannot power dominate more than one of its
out-neighbors due to rule (D2). Now, we need to prove that there is
no dependency cycle. By way of contradiction, suppose that
$C^*=u_1,u_2,\ldots,u_m$ is a dependency cycle. Focus on the edges
of $C^*$. Call the direction of the red edges \emph{forward}, and
call the direction of the blue edges \emph{backward}. Observe that a
dependency cycle has $\geq 1$ red edges, and each of its red edges
corresponds to an application of rule (D2). Assume that  all the
edges in $C^*$ are red. Then the red edges $(u_i,u_{i+1})$ imply
that
 $u_{i}< u_{i+1}$ for all
$i=1,2,\ldots,m-1$; therefore $u_1<u_2<\cdots<u_m$, but this is a
contradiction since the last red edge from $u_m$ back to $u_1$
implies that $u_m< u_1$. Hence, there is no dependency cycle with
all edges colored red. Now, assume that the dependency cycle $C^*$
has some blue edges.
 We show that a similar contradiction occurs when there are no two consecutive blue
edges. Consider a blue edge $(u,v)$ of $C^*$ and note that the other
edge of $C^*$ incident to $u$ is a red edge, say $(u,w)$. By rule
(D2), we see that $v$ should be power dominated before $u$ can power
dominate $w$; thus we have $w>v$. Repeating this argument, we get an
ordering for the occurrences of power domination of some of the
nodes in $C^*$ that gives a contradiction, e.g., if $m$ is even and
the edges of $C^*$ are alternatively blue and red (starting with
blue) we get $u_1<u_3<u_5<\cdots<u_{m-1}<u_1$ (see Figure
\ref{Fig:VldColrP} for an example). Hence, $G$ has a valid coloring
with $S$ as the set of origins.
\begin{figure}[!h]
\begin{center}
\input{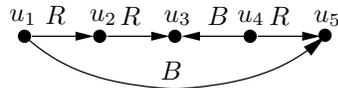}
\caption{ The blue edge $(u_4,u_3)$ means that $u_3$ should be power
dominated before we can power dominate $u_5$ by applying rule (D2)
on $u_4$; thus $u_3< u_5$. Again the blue edge from $u_1$ to $u_5$
implies that $u_5<u_2$. Finally , the red edge $(u_2,u_3)$ shows
that $u_2 <u_3$. Combining these dependencies we get
$u_3<u_5<u_2<u_3$. This is a contradiction, and shows that there
cannot be a dependency cycle  in a coloring obtained from the
application of rules (D1) and (D2) of \textsc{Directed PDS}.}
\label{Fig:VldColrP}
\end{center}
\end{figure}

Now suppose that $G$ has a valid coloring ${\mathcal{C}=(V,E_r\cup
E_b)}$ with $S\subseteq V(G)$ as the set of origins. The nodes in
$S$ and all of their out-neighbors in $G_r=(V,E_r)$  are power
dominated by applying the rule (D1). Now we prove that $S$ will
power dominate all nodes in $G$. Suppose  that this does not happen.
Let $X\subset V$ be the maximal set of nodes that can be power
dominated by $S$. We claim that there is at least one red edge from
$X$ to $V\setminus X$. Note that all of the origins are in $X$, so
each of the nodes in $V\setminus X$ has in-degree $1$ in $G_r$.
Hence, if there is no red edge from $X$ to $V\setminus X$, then
there should be a directed cycle of red edges in $G[V\setminus X]$.
This is not possible since there are no dependency cycles. Therefore
there is at least one red edge from $X$ to $V\setminus X$. Let
$e_1=(x_1,y_1),\ldots, e_k=(x_k,y_k)$ be all of the red edges from
$X$ to $V\setminus X$. If some $x_i$ has all of its out-neighbors in
$X$ except $y_i$, then by applying rule (D2) on $x_i$ the node $y_i$
will be power dominated. By the maximality assumption of $X$ this
cannot happen. Therefore, each $x_i$ has another out-neighbor, say
$z_i$, in $V\setminus X$. Then $(x_i,z_i)$ is a blue edge,
otherwise, $x_i$ would be an origin and $y_i$ would be power
dominated by applying rule (D1) on $x_i$. Now, we construct a
dependency cycle as follows: starting from $x_1$, use a blue edge to
move to a node $z_1$ in $V\setminus X$; then move in the reverse
direction over a sequence of red edges $(z_2,z_1), (z_3,z_2),
\cdots$ until we reach a red edge $(z_i,z_{i-1})$ with $z_{i-1}\in
V\setminus X$ and $z_i\in X$ (such an edge exists since
$G[V\setminus X]$ has no directed cycle of red edges); note that
$(z_i,z_{i-1})$ is one of the red edges $(x_1,y_1), \cdots,
(x_k,y_k)$. If $z_{i}=x_1$, then we have a dependency cycle;
otherwise, we again use a blue edge $(z_i,z_{i+1})$ to move to a
node in $V\setminus X$. By repeating these steps, we will eventually
find a dependency cycle. Note that all the blue edges are in one
direction, and all the red edges are in the other direction. This is
a contradiction, since $\mathcal{C}$ has no dependency cycle. Hence,
we have $X=V$, so $S$ power dominates $G$.
\end{proof}

\begin{theorem}
\label{Thm:DynProg} Given a directed graph $G$ and a tree
decomposition of width $k$ of its underlying undirected graph,
\textsc{Directed PDS} can be optimally solved in $O(c^{k^2}\cdot n)$
time for a global constant $c$.
\end{theorem}

A consequence of the above theorem is a linear-time algorithm for
solving the \textsc{Directed PDS} problem optimally on directed
graphs, given a bounded tree-width decomposition of the underlying
undirected graph. Also since the tree-width decomposition for graphs
with bounded tree-width can be computed in polynomial-time
\cite{Bib:BOD96}, there is a polynomial-time algorithm to solve
\textsc{Directed PDS} optimally on the class of directed graphs such
that the underlying undirected graph has  bounded tree-width.
\section{Conclusions}
We studied the PDS problem from the perspective of approximation
algorithms. We introduced a natural extension of the problem to
directed graphs. We showed that both problems have a threshold of
$O(2^{\log{n}^{1-\epsilon}})$ for the hardness of approximation. We
presented an $O(\sqrt{n})$ approximation algorithm for
\textsc{Planar PDS}. We designed a dynamic-programming algorithm for
solving the \textsc{Directed PDS} problem optimally in linear-time
for those directed graphs whose underlying undirected graph has
bounded tree-width.

Here, we describe an algorithm with an approximation guarantee of
$O(\frac{n}{\log{n}})$ for the PDS problem. The algorithm works as
follows. Partition the nodes of the graph $G$ into $\log{n}$
equal-sized sets $V_1, V_2, \cdots$. Next, consider all possible
ways of picking these sets (we pick all nodes in a set). Among all
these different candidates, output the one that power dominates $G$
and has the minimum number of nodes. Note that in the algorithm we
only consider $2^{\log{n}}=n$ different candidates. Clearly, the
algorithm runs in polynomial time, since the feasibility of each
candidate can be tested in polynomial time. Let $S^*$ be an optimal
solution. It is easy to see that the set of $V_i$'s that intersect
$S^*$ forms a feasible solution for the PDS problem in $G$; this
solution has size at most $\frac{n}{\log{n}}\cdot\card{S^*}$. This
establishes the approximation guarantee. The same algorithm and
analysis applies to the \textsc{Directed PDS} problem.
\begin{proposition}
There is a polynomial time $\frac{n}{\log{n}}$-approximation
algorithm for both the PDS problem and the \textsc{Directed PDS}
problem.
\end{proposition}
There is a gap between our hardness threshold of
$O(2^{\log{n}^{1-\epsilon}})$ and our approximation guarantee of
$O(\frac{n}{\log{n}})$, and narrowing this gap is an open question.

A major open question in the area is whether there exists a PTAS
({\em polynomial time approximation scheme}) for \textsc{Planar
PDS}. A first step may be to obtain an improvement on our
approximation guarantee of $O(\sqrt{n})$. There has been a lot of
research on designing  PTASs for $\np$-hard problems on planar
graphs. Some of the most important developments are the outerplanar
layering technique by Baker \cite{Bib:BA94}, and the
bidimensionality theory by Demaine and Hajiaghayi \cite{Bib:DEHA05}.
Unfortunately, these methods do not apply to \textsc{Planar PDS}.

Baker \cite{Bib:BA94} showed that the \textsc{Dominating Set}
problem in planar graphs has a PTAS. In the Baker method we first
partition the graph into smaller graphs. Then we solve the problem
optimally on each subgraph, and finally we return the union of the
solutions as a solution for the original graph. The example in
Figure \ref{Fig:Baker} shows that this method does not apply to
\textsc{Planar PDS}. The size of an optimal solution is $1$, but if
we apply the Baker method, then the size of the output solution will
be at least as large as the number of subgraphs in the partition
which can be $\Theta(n)$.

Demaine and Hajiaghayi \cite{Bib:DEHA05} introduced the
bidimensionality theory and used it to obtain PTASs for several
variants of the \textsc{Dominating Set} problem on planar graphs. An
important property of bidimensionality is that when an edge is
contracted the size of an optimal solution should not increase.
Consider the example in Figure \ref{Fig:BadBID}. If we contract
edges $e_1,e_2,\ldots,e_n$ in $G$, then we get the graph $G'$. It
can be checked that $\Opt(G)=1$ but $\Opt(G')=\Theta(n)$. Thus the
bidimensionality theory does not apply to \textsc{Planar PDS}, since
the optimum value may increase when an edge is contracted.
\begin{figure}[htbp]
\begin{center}
\input{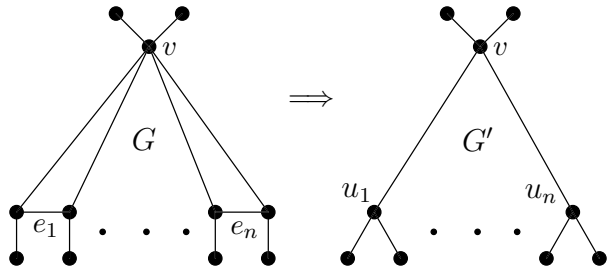}
\caption{ Optimal value of PDS increases when edges are
contracted.}\label{Fig:BadBID}
\end{center}
\end{figure}

Lastly, we consider some variations of greedy algorithms for PDS and
show that they perform very poorly even on planar graphs. In
contrast, for other related problems such as \textsc{Dominating Set}
and \textsc{Set Covering}, greedy algorithms perform well since they
achieve a logarithmic approximation guarantee, and no substantial
improvement is possible by any polynomial time algorithm, under
complexity assumptions like $\p\neq\np$. The most natural greedy
algorithm for PDS is the one that starts with $S=\emptyset$, and in
each step, adds a new node $v$ to the current solution $S$ such that
$v$ power dominates the maximum number of new nodes.

Unfortunately, this greedy algorithm  may find a solution $S$ such
that $\card{S}\geq\Theta(n)\cdot \Opt(G).$ To see this, consider a
graph $G$ that is obtained from a $9\ell\times 9m$ grid by
subdividing all row-edges, except with minor changes in the four
corners as shown in Figure \ref{Fig:Grid}. Partition the graph $G$
into  $9\times 9$ grids (ignoring the nodes introduced by
subdivision), see Figure \ref{Fig:Grid99}. It is easy to check that
any single node can power dominate at most $7$ nodes, and the center
node of any one of the $9\times 9$ grids achieves this maximum. So
the greedy algorithm at the first iteration may pick the center node
of any one of the $9\times 9$ grids.  Assuming all nodes picked by
the algorithm so far have been these center nodes, we see that
picking another center node maximizes $\left|\pd_{S\cup\set{
v}}\setminus \pd_{S}\right|$ over all $v\in V$. So the greedy
algorithm could continue picking center nodes, and after that
possibly picking other nodes until it finds a feasible solution $S$.
\begin{figure}[!h]
\begin{center}
\subfigure[Grid] {
    \label{Fig:Grid}
    \includegraphics{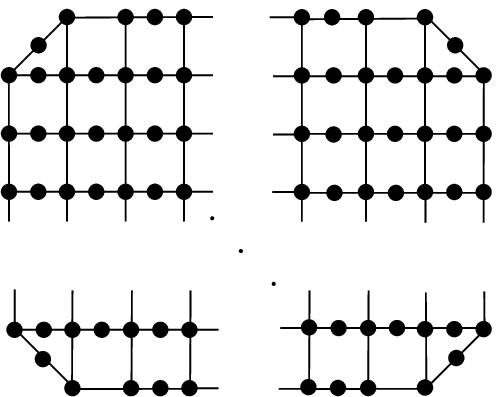}
} \hspace{1cm} \subfigure[$9\times 9$ grid] {
    \label{Fig:Grid99}
    \includegraphics{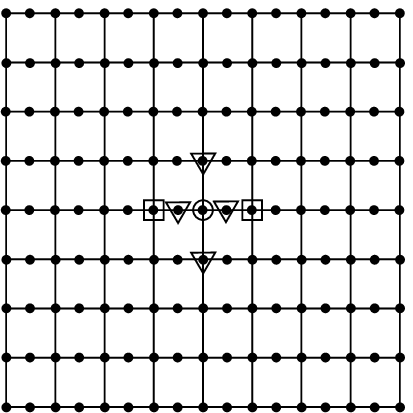}
} \caption{Bad example for the greedy algorithm} \label{Fig:GridSub}
\end{center}
\end{figure}
The size of the output $S$ is at least $m\cdot\ell=\Theta(n)$. By
Proposition \ref{Pro:GridOPT}, we have $\Opt(G)=\Theta(\ell)$. Now
by fixing $\ell=\Theta(1)$ we can see that the size of the output
solution can be bigger than $\Opt(G)$ by a factor of $\Theta(n)$.
\begin{proposition} The greedy algorithm for the PDS problem may find a
solution $S$ such that $\card{S}\geq\Theta(n)\cdot \Opt(G).$
\end{proposition}

We will consider  two other variations of the greedy algorithm,
namely \emph{Proximity} and \emph{Cleanup}.   We have examples of
\textsc{Planar PDS} showing that these variations of the greedy
algorithm perform poorly.

\noindent{\bf Proximity:} In each step of the \emph{Proximity}
algorithm we choose a node such that the set of all power dominated
nodes induces a connected subgraph, and subject to this, the number
of newly power dominated nodes is maximized. Informally, this is to
escalate the use of the propagation rule.

The bad example for the proximity version of the greedy algorithm is
obtained by modifying the center row of the $h\times(2m+1)$ grid, as
shown in Figure \ref{Fig:GrdyProx}, by inserting $\ell$ subdividing
nodes into the edges in the middle row, and also subdividing all of
the other row-edges except some of the corner edges. Figure
\ref{Fig:GrdyProx} illustrates an example of such a grid for $l=5$
and $h=9$ rows, but for a bad example for the proximity greedy
algorithm we need $h$ to be sufficiently large constant ($h=17$
suffices). We use the figure for illustration, to show the working
of the proximity greedy algorithm.

It is easy to check that by picking all nodes of the first column we
can power dominate the entire graph, so the optimal solution is
$\Theta(1)$. The proximity greedy algorithm starts by picking a node
that power dominates maximum number of nodes (which is
$17=2\ell+7$); any white node satisfies this requirement. Therefore
the algorithm may pick for example the first white node (from the
left). It is easy to check that in the next step the algorithm will
pick the white node to the right of the first one, since all of the
power dominated region stays connected and also it power dominates
maximum number of new nodes (which is $16=2\ell+6$). The algorithm
continues picking all white nodes and at the end it will pick
possibly more nodes to get a feasible solution. (The shaded region
shown in Figure \ref{Fig:GrdyProx} indicates the nodes that will be
power dominated by picking all white nodes.) Therefore, the size of
the solution found by the algorithm is at least $m=\Theta(n)$.
Hence, the proximity greedy algorithm may find a feasible solution
that is $\Theta(n)$ times worse than the optimal solution.
\begin{figure}[!h]
\begin{center}
    \includegraphics[scale=1]{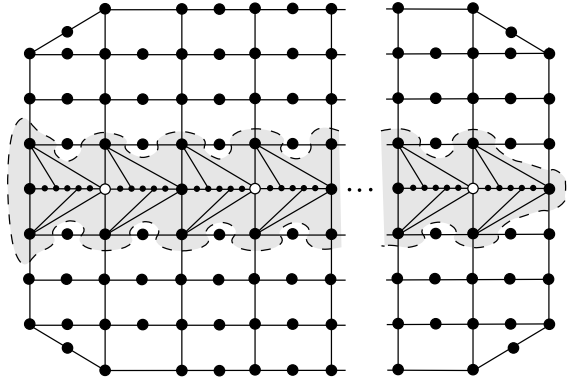}
\caption{Bad example for the proximity greedy algorithm}
\label{Fig:GrdyProx}
\end{center}
\end{figure}

\noindent{\bf Cleanup Step:} Some of the recent approximation
algorithms, especially some based on the primal-dual method, use a
clean up step at the end: this step removes redundant elements from
the solution in some sequential order. In the \emph{Cleanup}
algorithm, we first run the greedy algorithm to find a solution
(node set) $S$, then we repeatedly remove nodes from $S$, until $S$
is an inclusionwise minimal power dominating set. Although a cleanup
step may substantially improve on the solution found by the greedy
algorithm on some examples, this does not hold for all examples.

The same bad example for the proximity version  is also a bad
example for the cleanup version of the greedy algorithm. The cleanup
greedy algorithm may again pick the first white node (from the
left), and after picking this white node, it may pick the third
white node. Since both of them power dominate maximum number of new
nodes (which is $17=2\ell+7$). Note that in the original greedy
algorithm there is no need to have a connected subgraph induced on
power dominated nodes. The algorithm continues to pick all the odd
indexed white nodes, and after that it will start picking the even
indexed white nodes since any one of them power dominates maximum
number of new nodes (which is at least $15=2\ell+5$). At this stage
of the algorithm the set of power dominated nodes are those in the
shaded region of Figure \ref{Fig:GrdyProx}. It is easy to check that
we need to pick nodes from both upper and lower parts in order to
power dominate the entire graph. The greedy algorithm may pick some
nodes from the leftmost column in the top part and some nodes from
the rightmost column in the bottom part to power dominate the entire
graph. Now we start doing the cleanup process. It can be checked
that if we remove any two consecutive white nodes from the obtained
solution, the graph cannot be power dominated completely. So we need
to keep at least half of the white nodes. Therefore, the size of the
output solution at the end of cleanup process is at least
$\frac{m}{2}=\Theta(n)$, but the optimal solution is just
$\Theta(1)$ as before.

%\input{Conc}

%\bibliography{Bib-arXiv}
\clearpage
\appendix

\section{Dynamic Programming}
\label{Apn:A} In this section we describe our dynamic-programming
algorithm for the \textsc{Directed PDS} problem. This algorithm is
similar to the dynamic-programming algorithm designed by Guo et
al.\cite{Bib:GUNR05} to optimally solve PDS for undirected graphs
with bounded tree-width. It is known  that any tree decomposition of
width-$k$ can be transformed to a nice tree decomposition with width
$k$ in linear-time \cite{Bib:KL94} (Lemma 13.1.3). So we can assume
that we are given a nice tree decomposition of the underlying
undirected graph of $G$ call it $\langle \set{X_i| i\in
I},T\rangle$. Let $T_i$ denote the subtree of $T$ rooted at $T$-node
$i$, and $Y_i=\left(\bigcup_{j\in V(T_i)}{X_j}\right)\setminus X_i$.
Also let $G_i$ be the subgraph induced by $Y_i\cup X_i$, i.e.\
$G_i=G\left[Y_i\cup X_i \right]$, and let ${G'}_i=G[X_i]$.

Consider a valid coloring $\mathcal{C}$  of the graph $G$. We store
the color of the edges in each bag by assigning a state to that bag
(the formal definition of a state will follow). We can reconstruct
the coloring $\mathcal{C}$ from the states of all bags in the tree
decomposition of $G$; so there is no  need to store the coloring
$\mathcal{C}$ in the dynamic-programming.

\noindent{\bf The state of a bag:} Given a coloring $\mathcal{C}$,
the state of a bag $X_i$ describes the coloring of the edges in
$G'_i$. In order to detect the dependency cycles in the coloring
$\mathcal{C}$ without reconstructing the whole coloring, we need to
store some more information in a state. This extra information
enables us to detect a dependency cycle in $G_i$ which goes through
$X_i$, by considering only the state of the bag $X_i$. A {\it bag
state} $s$ contains the following: state of each edge, state of each
node, and state of each pair of nodes in $G'_i=G[X_i]$.
\begin{itemize}
\item  \texttt{State of an edge}: The state of an edge $e\in E(G[X_i])$
denoted by $s(e)$ is the color that is assigned to $e$ in the
coloring $\mathcal{C}$; $s(e)\in\set{R,B}$.
\item  \texttt{State of a node}: The state of a node $v\in X_i$
denoted by $s(v)$ shows the number of red edges between $v$ and
$Y_i$.
\begin{itemize}
\item $s(v)=1$: There is exactly one red edge from a node in $Y_i$
to $v$ and no red edge from $v$ to $Y_i$,
\item $s(v)=2$: There is exactly one red edge from a node in $Y_i$ to
$v$ and exactly one red edge from $v$ to $Y_i$,
\item $s(v)=3$: There is no red edge between $Y_i$ and $v$,
\item $s(v)=4$: There are at least two red edges from $v$ to $Y_i$
and no red edge from $Y_i$ to $v$,
\item $s(v)=5$: There is exactly one red edge from $v$ to $Y_i$ and
no red edge from $Y_i$ to $v$.
\end{itemize}
\item  \texttt{State of a pair of nodes}:  A {\it dependency path} from $u$ to $v$ is a path
$P$ where all red edges in $P$ are directed from $u$ to $v$ and all
blue edges are directed from $v$ to $u$. We categorize dependency
paths according to the color of their first and last edges. There
are $4$ possible types $RR,RB,BR,BB$; for example a path of type
$RB$ is a path with the first edge colored red and the last edge
colored blue. For a pair $(u,v)\in X_i\times X_i$ ($u\neq v$) the
state of $(u,v)$ denoted by $s(u,v)$ shows the type of dependency
paths from $u$ to $v$ in $G[Y_i\cup\set{u,v}]$; that is,
$s(u,v)\subseteq\set{RR,RB,BR,BB}$. Note that there are $2^4=16$
different states for each pair of nodes.
\end{itemize}

{\noindent \bf Detecting dependency cycles:}: An important part of
the dynamic-programming algorithm is to detect dependency cycles in
the coloring $\mathcal{C}$. Assume we are at bag $X_i$ and we are
given the bag state $s$ corresponding to the coloring $\mathcal{C}$.
We can detect the dependency cycles in $G'_i=G[X_i]$ by enumerating
all possible cycles; note that the coloring of edges in $G'_i$ is
given in the state $s$. The dependency cycles in $G_i$ can be
detected by considering the state of each pair of nodes in $X_i$.
For example assume that $RB\in s(u,v)$ and $RR\in s(v,u)$. Then, by
combining a dependency path of type $RB$ from $u$ to $v$ and a
dependency path of type $RR$ from $v$ to $u$ we obtain a dependency
cycle going through $u$ and $v$ in $G_i$.

Let us denote by $\Lambda_i$ the set of all possible states for the
bag $X_i$. The dynamic-programming will compute a mapping
$A_i:\Lambda_i\rightarrow\mathbb{N}\cup\set{+\infty}$. For a bag
state $s\in\Lambda_i$ the value  $A_i(s)$ is the minimum number of
origins in an optimal valid coloring $\mathcal{C}$ of $G_i$ under
the restriction that the state of nodes, edges, and pairs of nodes
in $X_i$  is given by $s$. Now, we describe how our
dynamic-programming works.

{\noindent \bf Step 1: (Initialization):} In this step for each leaf
node $i$ of $T$, we initialize the mapping $A_i$ as follows. For a
given state $s$, we define $A_i(s)$ as $+\infty$ if $s$ has a
dependency cycle, a node $v$ with $s(v)\neq 3$, or a pair of nodes
$u$ and $v$ such that $s(u,v)\neq\emptyset$. Otherwise, we define
$A_i(s)$ as the number of nodes with no in-coming red edges in the
coloring defined by $s$.

{\noindent \bf Step 2: (Bottom-Up Computation):} After
initialization, we visit the nodes in $T$ in a bottom-up fashion and
at each bag $X_i$ we compute the mapping $A_i$ corresponding to
$X_i$. The update process depends on the type of $T$-nodes that we
are considering. Here, we only consider the update process at an
\textsc{Insert Node}. The other cases are similar to this one.

{\noindent \bf Insert Node:} Suppose $i$ is an insert node with the
child $j$, and assume that $X_i=X_j\cup\set{x}$. For each bag state
$s\in\Lambda_i$ do the following:
\begin{enumerate}
\item Check whether the coloring given by $s$ forms a
valid coloring of $G'_i$; if not, define $A_i(s)=+\infty$.
\item Compute the set $\Lambda_j(s)$ containing bag states of $j$ that
are ``compatible'' with the bag state $s$.
\item For each $s'\in \Lambda_j(s)$, check if
a valid coloring of $G_j$ ``compatible'' with $s'$ can be extended
to a valid coloring of $G_i$ ``compatible'' with $s$.
\item Compute $A_i$ based on the mapping $A_j$.
\end{enumerate}
{\noindent {\bf Compatible bag state} (Step 2):} A bag state
$s'\in\Lambda_j$ is said to be \emph{compatible} with the bag state
$s\in\Lambda_i$ if the state of each node, each edge, and each pair
of nodes in $V(G'_j)$ in the bag state $s'$ is the same as  the
corresponding state in the bag state $s$. If $s(x)\neq 3$, or
$\exists v\in X_j:s(x,v)\neq \emptyset \vee s(v,x)\neq \emptyset$
then we define $\Lambda_j(s)=\emptyset$.

{\noindent {\bf Detecting dependency cycles} (Step 3):} The
conditions of a valid coloring can be violated due to degree
constraints on the new node $x$, or by the existence of a dependency
cycle going through $x$. Both these cases can be tested by
considering the bag states $s$ and $s'$.

{\noindent {\bf Computing $A_i$} (Step 4):} The addition of $x$ may
change the number of origins in $G_i$. The node $x$ will be an
origin if it has at least one outgoing red edge in $s$. But an
origin node $v\in X_j$ (in $s'$) that has an incoming red edge from
$x$, is no longer an origin. So by considering the red edges going
out of $x$ we can update the number of origins and compute the
mapping $A_i$. If the coloring compatible with $s$ is not a valid
coloring, then  we define $A_i(s)$ to be $+\infty$.

{\noindent \bf Step 3: (At root $r$)}: Finally, we compute the
number of origins in an optimal valid coloring of $G$ by finding the
minimum of $A_r(s)$ over all possible $s\in\Lambda_r$. It is easy to
see that each bag ${X_i}$ has at most $16^{(k+1)^2}\cdot
5^{k+1}\cdot 2^{(k+1)^2}$ states; note that $\card{X_i}\leq (k+1)$.
It can be checked that the total running time of our algorithm is
 $O(c^{k^2}\cdot n)$, for some global constant $c$.
\end{document}